\documentclass[aps,prl,twocolumn,groupedaddress]{revtex4}
\usepackage{graphicx}
\usepackage{amsmath}
\usepackage{subfigure}

\begin{document}
\title{Spin Valve Effect in ZigZag Graphene Nanoribbons by Defect Engineering}
\author{Sankaran Lakshmi$^{1}$, Stephan Roche$^{1,2}$ and Gianaurelio Cuniberti$^{1}$}
\affiliation{$^{1}$ Institute for Materials Science and Max Bergmann Center of Biomaterials,\\
Dresden University of Technology, D-01062 Dresden, Germany\\ $^{2}$ CEA, 
Institute for Nanoscience and Cryogenics, INAC/SP2M/${\rm L}_{\rm sim}$, 17 rue des Martyrs, 
38054 Grenoble Cedex 9, France}

\begin{abstract}
We report on the possibility for a spin valve effect driven by edge defect engineering of zigzag 
graphene nanoribbons. Based on a mean-field spin unrestricted Hubbard model, electronic 
band structures and conductance profiles are derived, using a self-consistent scheme to include 
gate-induced charge density. The use of an external gate is found to trigger a semiconductor-metal 
transition in clean zigzag graphene nanoribbons, whereas it yields a closure of the spin-split 
bandgap in the presence of Klein edge defects. These features could be exploited to make novel 
charge and spin based switches and field effect devices.

\end{abstract}
\pacs{81.05.Uw,75.75.+a,72.25.-b}
\maketitle
\date{\today}

The recent fabrication of single and few layers graphene has sparked considerable expectations 
in the field of all-carbon nanoelectronics \cite{graphene,GeimRMP}. Although 2D graphene is a 
zero gap semiconductor, the possibility to pattern graphene nanoribbons (GNRs) 
with widths of few tens of {\it nm}, has enabled band gap engineering \cite{Kim} and the 
development of efficient GNRs-based field effect transistors \cite{GNRFET}. However, the 
further optimization of GNRs-based devices or the design of novel device functionalities 
demand for more understanding and control of not only unavoidable structural disorder and 
defects, but also edge geometry and chemical functionalization. Experimental characterization 
either with Raman studies, scanning tunneling microscopy or high-resolution electronic
transmission microscopy has revealed a large spectrum of topologically different edge profiles, 
which can exhibit either an armchair or a zigzag symmetry \cite{GNRzza}, as well as a zoology of 
edge imperfections \cite{Edgeprofiles}. In particular, individual pending atoms named Klein defects 
have been predicted \cite{Klein1} and more recently observed \cite{Klein2,Klein3}. In principle, such 
edge disorder is unsuitable for keeping good conduction properties of otherwise clean GNRs. As a 
matter of fact, several experimental \cite{Stampfer09} and theoretical \cite{White} reports evidence the 
enlargement of the transport gap and strong fluctuations of low temperature conductance. 

A pioneering work by Son {\it et al.} \cite{Louie} has unveiled half-metallic behavior in clean
ZGNRs in the presence of a transverse electric field, opening perspectives
for the design of spin-dependent switching devices \cite{Yazyev}. Several recent studies have 
also reported on the possibility for half-metallicity in ZGNRs using chemical doping \cite{Dutta}, 
whereas the application of an external magnetic field was shown to trigger a transition from 
parallel to antiparallel magnetic edges, resulting in giant magnetoresistance phenomena and 
efficient spintronics devices \cite{Palacios}.

In this Letter, we investigate the electronic and conductance properties of gated zigzag GNRs (ZGNRs) 
with and without the presence of Klein edge defects. The energetics of the GNRs is described by a 
mean-field Hubbard Hamiltonian which has been shown to well reproduce density functional 
theory results in the spin local density approximation \cite{Palacios1,Palacios2}. The gate-injected charge 
density is also self consistently included in the band structures and transport calculations. 
For clean zigzag GNRs, the external gate is shown to tune the electronic structure from a 
semiconducting state to a purely metallic state, thus switching on current through 
the device, whereas in presence of Klein defects, the spin-split gap is closed, resulting in a 
transition from a pure spin-current to a spin-degenerate charge current, close to the Fermi energy.

ZGNRs are very peculiar in the sense that they have a magnetic ground state, with each edge having electrons 
aligned ferromagnetically with each other, albeit anti-ferromagnetically with respect to the other 
edge \cite{Louie}. The ground state is entirely dominated by these magnetic edge states 
\cite{GNRzza}, which generate a semiconducting band gap \cite{Louie}. The introduction of a Klein edge 
profile (with additional $\pi$-electron hopping bonds) in the ZGNR, is known to produce a flat band 
over the entire Brillouin zone \cite{Klein2}, within a tight-binding description. 
However, with the incorporation of electron correlations, a ground state with a net 
magnetization and a spin-split band gap emerges \cite{Lieb,Kusakabe}.
Kusakabe {\it et al.} have theoretically discussed a Klein edged situation 
by dihydrogenating one of the edges of the ZGNR, validating the occurence of a spin-polarized band 
at the Fermi energy \cite{Kusakabe}. Very recently, these edges were observed for the first time, 
although locally, in graphene using HR-TEM \cite{Klein3}.

Here, it is our interest to study the effect of an external gate on the band gaps of these
two categories of bipartite semiconducting systems, one of which has a anti-ferromagnetic ground 
state (ZGNR), and the other ferromagnetic (ZGNR with a Klein edge defect). We model GNRs using the 
spin-unrestricted mean-field Hubbard Hamiltonian,

\begin{equation}
H = -t \sum\limits_{\langle ij\rangle,\sigma} (c_{i\sigma}^\dagger c_{j\sigma}+\textrm{h.c}) + U \sum\limits_{i,\sigma}\langle n_{i-\sigma}\rangle n_{i\sigma}
\end{equation}

\begin{figure*}
\centering
\includegraphics[scale=0.5]{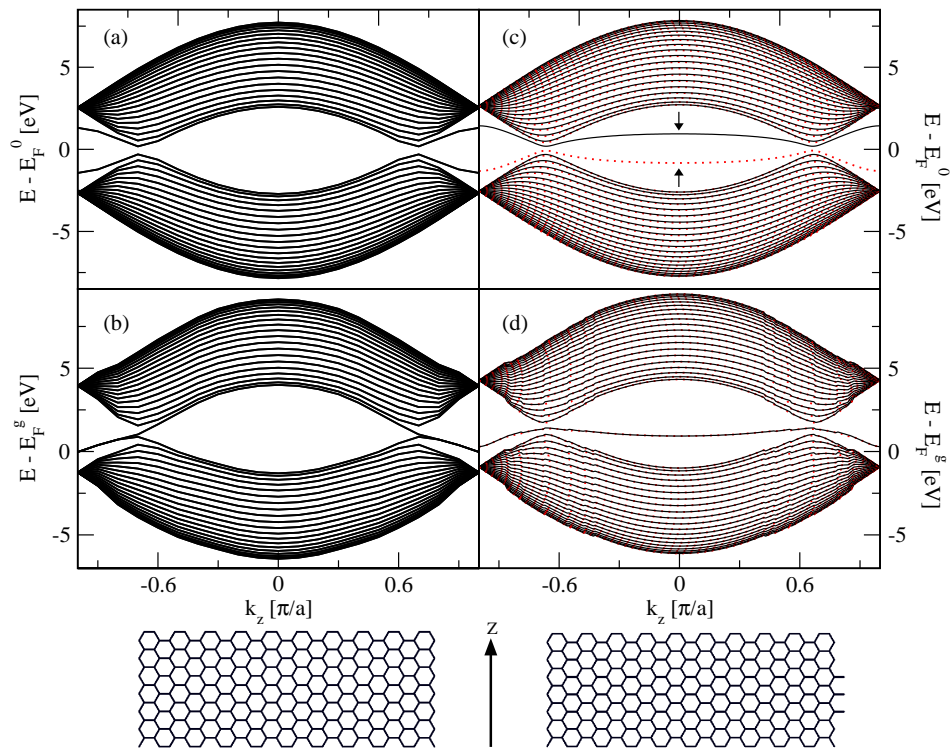}
\caption{(color online) Band structures of the clean ZGNR with (a) $V_g=0$ (b) $V_g=1.5$ V for both up
and down (degenerate) spins. Band structures of the ZGNR with one Klein edge when (c) $V_g=0$ 
(d) $V_g=1.5$ V for up (red dotted) and down (black solid) spins. The schematic shows few unit cells of the
clean ZGNR on the left. On the right, three Klein defect edges in an otherwise clean ZGNR have 
been shown for clarity.\\}
\end{figure*}

\noindent where $c_{i\sigma}^\dagger$, $c_{i\sigma}$ and $n_{i\sigma}$ are the creation, annihilation and
number operators for an electron of spin $\sigma$ in the $\pi$ orbital of the $i^{\rm th}$ C atom in the ribbon,
with $t=2.6$ eV and $U=2.75$ eV \cite{White}. The unit cell of the ZGNR has width 20 (40 atoms)
and that of the Klein edged ZGNR has an extra C atom at one edge.
The source-drain electrodes, for simplicity are assumed to consist of the same GNRs. The Fermi energy of the
non-gated system ($E_F^0$) is obtained by integrating the charge density up to half-filling. 

One common way to tune the device conductance is to use an external metallic gate for monitoring the 
depletion or accumulation of charges in the conducting channel \cite{GNRFET}. The presence of a third (gate) terminal is numerically simulated by shifting the on-site energies by $eV_g$ where $V_g$ is the gate voltage. The modified charge density distribution up to $E_F^0$ at every $V_g$ is further self-consistently computed. To do so, 
the excess charge carriers introduced by the gate are obtained as $n(E_F^g)-n(E_F^0)$ where $n(E_F^g)$ is 
the charge density at the Fermi energy in the presence of the gate and $n(E_F^0)$ is that in the absence of the gate
 ($n(E_F)=\int^{E_F}_{-\infty} \rho(E) \, dE$, with $\rho(E)$ the density of states). The modified charge density at every $V_g$ is then incoporated into the Hamiltonian 
and solved self-consistently, to obtain the new band-stucture of the gated GNR, with the associated Fermi level ($E_F^g$).

The coherent transport through the system is then calculated, using the well-known Landauer's formalism.
The retarded Green function of the device is computed as $G=(E - H -\Sigma_L - \Sigma_R)^{-1}$
where $\Sigma_L$ and $\Sigma_R$ are the self-energies of the GNR electrodes, calculated as
$\Sigma=\tau g_s \tau^{\dagger}$ where $\tau$ contains the device-source/drain interactions and $g_s$ is the
surface Green's function of the electrodes, calculated using standard recursive Green's function
techniques \cite{White}. The conductance of the device is finally obtained as
$G_D(E) = \frac{e^2}{h} T(E)$ where T(E), the transmission probability at energy E is given by
 ${\rm Tr}[\Gamma_L G \Gamma_R G^{\dagger}]$,
Tr is the Trace and $\Gamma_{L,R}=i(\Sigma_{L,R}-\Sigma_{L,R}^{\dagger})$.

Electronic band structures of ZGNR with or without a Klein edge, and as a function of applied external gates 
can be seen in Fig 1. First, as observed in Fig 1a, when $V_g=0$,
the ground state is characterized by a band gap, resulting from its oppositely spin-polarized edges.
With the application of a positive gate voltage channeling holes into the system, the electron
density at the edges ($2/3 \leq \lvert k \rvert \leq 1 $) gradually diminishes, which results in a completely
non-magnetic, metallic system. A closing of the band gap is seen in Fig 1b for
$V_g=1.5$ V. The effect of gate voltage on the low-energy conductance can be appreciated in Fig 2. Given that the GNRs 
are disorder free, the conductance remain quantized and is fixed by the number of available conduction channels 
at the relevant Fermi level. At injection energies close to $E_F^g$, a current switch is driven by the gate voltage 
increase, indicating a transistor type behavior.

In contrast, the situation for the ZGNR with one Klein edge shows some marked differences. First by looking at the band structure at zero gate voltage (Fig.1c), it is clear that the highest occupied band is made up entirely of one spin (dotted) and the lowest unoccupied band, of the other (solid). The ground state of the system shows some ferromagnetism, with both edges displaying high electron density of the same (majority) spin. This is explained by the finite sublattice imbalance which favors the appearance of midgap states and magnetic properties \cite{Palacios1,Palacios2,Inui94}.

When a positive gate voltage is applied, the majority spin (up) band shifts upward, reducing the electron (up spin) density, as seen in Fig 3b, which eventually meets the minority spin (down) density to make the system completely non-magnetic and metallic. This is also characterized in Fig 3a by the slow reduction of the spin-split gap, which eventually disappears beyond $V_g=1.5$ V.
The bandstucture of the GNR at $V_g=1.5$ V, shown in Fig 1d, clearly exhibits spin-degenerate
bands. When a negative gate bias injecting excess electrons into the device is applied, a similar situation follows.
The down spin band gets pushed into the occupied states, diminishing the total magnetization and hence the
spin-split gap. Here, the down spin density increases to meet its up spin counterpart as seen in Fig 3b.

\begin{figure}
\centering
\includegraphics[scale=0.32]{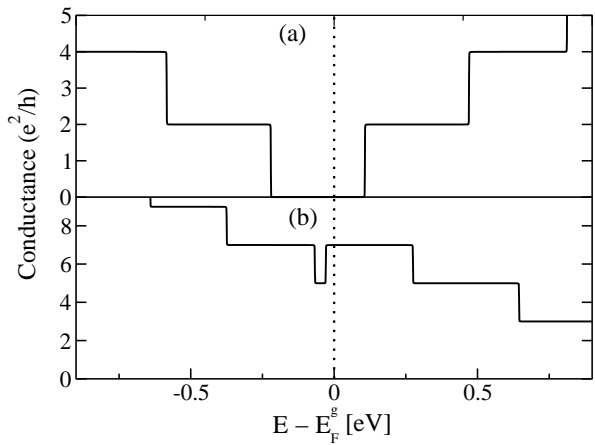}
\caption{ Spin-degenerate (up and down) conductance of the ZGNR with respect to the incident energy
of the electrons (E) scaled by $E_F^g$ for (a) $V_g=0$ (b) $V_g=1.5$ V.}
\end{figure}

In Fig. 4 at $V_g=0$, the transmission probability of the majority spin
close to $E_F^g$ is very high, as it is the only band available in the energy range. As the gate voltage
is increased, the system transmits only the majority spin until $\approx 1.0$ V, after which it turns
non-magnetic with both spins having identical transmissions. This indicates that it is possible to tune the current
in the system from one which is completely spin-polarized to an unpolarized charge current just with the
help of an external gate.

The origin for such transitions from semi-conducting to metallic in the gated ZGNRs and from a spin-polarized
to a spin-degenerate response in the Klein edged ZGNRs can be rationalized as follows. The ground state of both systems are
entirely composed of magnetic edge states, which produce their respective band gaps. An external gate voltage does break the spin polarization of the system
due to the change of the charge number at the edges. As a consequence, a controlled switch from a spin-splitting gap to metallicity is observed. One notes that the application of an external magnetic field in clean ZGNR also yields a similar effect and giant magentoresistance \cite{Palacios2}.

In summary, we have studied the effect of gate voltage on the band-structure and conductance profiles of a ZGNR and
a Klein edged ZGNR, based on the mean-field spin-unrestricted Hubbard model. The gate-injected charge density
was obtained self-consistently, and conductance calculations were performed using the Landauer's formalism.
We have found that an external gate could tune the spin-induced band-gap in clean ZGNRs or in Klein edged ZGNR.
In the former case, this leads to a switching-on of the charge current, whereas in the latter case, a transition from
a pure spin current to a completely unpolarized charge current is achieved. These features could eventually help in designing
 GNR based charge and spin switches and FETs in future all-carbon circuits.

\begin{figure}
\centering
\includegraphics[scale=0.3]{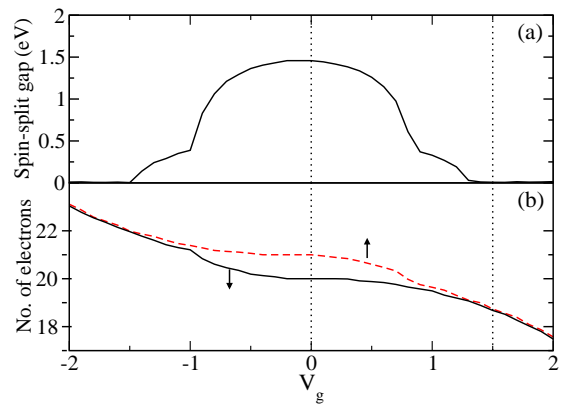}
\caption{(color online) (a) Variation of the bandgap over the entire Brillouin zone with $V_g$
(b) The total number of up (red dashed) and down electrons (black solid) with respect to $V_g$.
Dotted lines are marked at $V_g=0$ and $V_g=1.5$ V for clarity.}
\end{figure}

\begin{figure}
\centering
\vspace{0.5cm}
\includegraphics[scale=0.32]{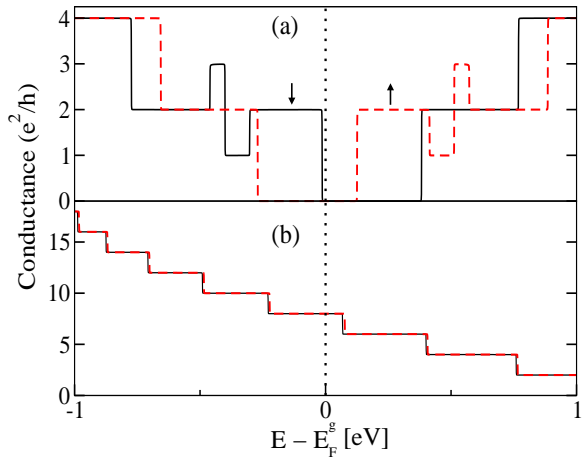}
\caption{(color online) Conductance of the Klein edged ZGNR with respect to the incident energy of the electrons ($E$)
scaled by $E_F^g$ for (a) $V_g=0$ (b) $V_g=1.5$ V. Dashed (red) lines refer to the spin-up conductance and solid
(black) lines refer to the spin-down conductance.}
\end{figure}

Support from the Alexander von Humboldt Foundation and computing facilities provided
by the ZIH at the Dresden University of Technology are duly acknowledged. This work was partially supported 
by the EU project CARDEQ under grant N0 IST-021285-2.

{}	

\end{document}